\documentclass{article}
\usepackage{amssymb}
\usepackage{amsfonts}
\usepackage{amsmath}

\input{tcilatex}

\begin{document}

\title{$1^{-+}\pi _{1}\left( 1400\right) $ cannot be hybrid meson}
\author{F. Iddir\thanks{%
E-mail: $iddir@univ$-$oran.dz$} and L. Semlala\thanks{%
E-mail: $l\_semlala@yahoo.fr$} \\
Laboratoire de Physique Th\'{e}orique, Univ. d'Oran Es-S\'{e}nia, ALGERIA}
\maketitle

\begin{abstract}
We expose arguments concerning the $\pi _{1}(1400)$, observed decaying into $%
\eta \pi $ and claimed to be an exotic meson. We conclude that this object
cannot be a $q\overline{q}g$ hybrid meson.
\end{abstract}

\section{Introduction}

From experimental efforts at IHEP, KEK, CERN and BNL, several isovector J$^{%
\text{PC}}$=1$^{\text{-+}}$exotic mesons have been identified. In this note
we focus on the resonance $\pi _{1}(1400)$ claimed to be decaying into $\eta
\pi $. More detailed data are summarized in table 1$^{\left[ 1\right] }$.
This discovery would be the first resonant state with really exotic quantum
numbers since no $q\overline{q}$ system can have J$^{\text{PC}}$=1$^{\text{-+%
}}$.

No glueball interpretation can be retained when the resonance has isospin 1.
Thus, we are left with two possibilities: the hybrid state $q\overline{q}g$
or the diquonium $q\overline{q}q\overline{q}$.

Hybrids have been studied, using the quark model with constituent gluon$^{%
\left[ 2-6\right] }$, the flux-tube model$^{\left[ 7\right] }$, using many
body Coulomb gauge Hamiltonian$^{\left[ 8\right] }$, QCD sum rules$^{\left[ 9%
\right] }$, the MIT bag model$^{\left[ 10\right] }$, and lattice-QCD$^{\left[
11\right] }$. These models predict that the lightest hybrid mesons will be J$%
^{\text{PC}}$=1$^{\text{-+}}$ meson, in 1.4-2.1 GeV mass range. Decay
process of the hybrid meson is studied in the framework of the first four
models.

Diquonium was the subject of several works$^{\left[ 3,12,13\right] }$.

\section{$\protect\pi _{1}(1400)$ cannot be hybrid meson}

Indeed, several arguments support this idea:

\ \ \ \ $i$) One difficulty with interpreting this state as a hybrid is the $%
\sim $500-600 MeV mass difference between the $\pi _{1}(1400)$ mass and flux
tube, LGT and MBCGH estimates of mass $\sim $1.9-2.1 GeV. In other hand, it
appears difficult to accommodate the $\pi _{1}(1400)$ as hybrid state in the
context of recent QCD sum rules calculations, the lightest hybrid in this
case being approximately with mass 1.6 GeV. Although the range mass values
1.3-1.7 obtained from bag models are believed to suffer from parameter
uncertainties$^{\left[ 14\right] }$.

\ \ \ $ii$) The studies of the decay mechanism$^{\left[ 7,9\right] }$
predict the suppression of \ 1$^{\text{-+}}$(hybrid) $\longrightarrow \eta
\pi $. This selection rule was generalized in a quite model independent way$%
^{\left[ 4\right] }$. In spite of our calculations of hybrid masses which
yield exotic 1$^{\text{-+}}$mesons with mass around 1.4 GeV (QE hybrid)$^{%
\left[ 6\right] }$ consistent with the experimental $\pi _{1}(1400)$ the
selection rule cited above excludes interpreting this resonance as QE-mode
hybrid.

\section{The $q\overline{q}q\overline{q}$ possibility}

The quantum numbers of $\pi _{1}(1400)$ states is rather typical for P-wave $%
q\overline{q}q\overline{q}$ states, so it is possible to interpret them as
light $q\overline{q}q\overline{q}$.

But here are some difficulties:

\ \ \ \ $i$) Mass results of 1$^{\text{-+}}$ $q\overline{q}q\overline{q}$%
.are around 1.7 GeV and Bag model does not predict the low-laying P-wave $q%
\overline{q}q\overline{q}$ states with I=1 as $\pi _{1}(1400)$.$^{\left[ 13%
\right] }$

\ \ $ii$) The selection rule prevents the decay of the P-wave $q\overline{q}q%
\overline{q}$ into $\eta \pi $.$^{\left[ 3\right] }$

\section{An alternative solution!}

Authors of [15] exploit the possibility to construct quark bilinear operator
which have J$^{\text{PC}}$=1$^{\text{-+}}$ exotic numbers. A numerical
solutions of Bethe-Salpeter equation yield two exotic 1$^{\text{-+}}$mesons
with masses of 1.439 and 1.498 GeV in good agreement with the experimental $%
\pi _{1}(1400)$. Their approach does introduce explicitly gluon degrees of
freedom. This may provide additional decay channels for exotic mesons, such
as $\pi _{1}(1400)\rightarrow \eta \pi $, that would normally be suppressed
according to OZI rule$^{\left[ 4\right] }$.

But the odd-time-parity amplitudes associated with the exotic mesons
corresponding to bound states whose BS amplitudes has negative
normalizations. The authors claim that this is not enough to disregard the
BS equation in these channels.

\section{Do $\protect\pi _{1}(1400)$ really an exotic resonance?}

In the work [16],\ since the attempts to describe the mass dependence of the
amplitude and phase motion with respect to the D wave as Breit-Wigner
resonance are problematic; the authors propose an alternative description of
the mass dependent P-wave amplitude and phase \textit{that does not require
the existence of an exotic meson }but is consistent with $\eta \pi $
re-scattering. In other hand, \textquotedblright there are possibility that
the structure observed by Crystal Barrel at 1400 is due to the rapid opening
of the threshold for the process $\eta \pi \longrightarrow b_{1}\pi $ (or $%
f_{1}\pi $), since that the threshold for $b_{1}($1235$)\pi $ is about 1370
MeV. On the Agrand diagram, the $\eta \pi $ amplitude will turn rapidly
though 90$^{\circ }$ if this channel opens, and will look as
resonance\textquotedblright $^{\left[ 17\right] }$ .

\section{Conclusion}

Taking into account the arguments cited above, $\pi _{1}(1400)$ ( if it has
really exotic quantum numbers) could not be\ an hybrid meson.

\bigskip

{\Large Aknowledgments}

We are grateful to D.V. Bugg for extremely useful discussions on the
experimental data.

\section{References}

[ 1] \ \ E. Klempt, arXiv: hep-ph/0404270v1

[ 2] \ \ A. LeYaouanc, L. Oliver, O. P\`{e}ne, J.-C. Raynal and S. Ono; Z.
Phys. Rev. \textbf{C28} \ (1985) 309

[\ 3] \ \ F. Iddir, A. LeYaouanc, L. Oliver, O. P\`{e}ne, J.-C. Raynal and
S. Ono, Phys.Lett.\textbf{B204}(1988)564

[ 4]\ \ \ F. Iddir, A. LeYaouanc, L. Oliver, O. P\`{e}ne, J.-C. Raynal ,
Phys.Lett.\textbf{B207}(1988)325

[ 5] \ \ F. Iddir and A.S. Safir, Phys.Lett.\textbf{B507}(2001)183, F. Iddir
and L. Semlala, arXiv: hep-ph/0411074

[ 6] \ \ F. Iddir and L. Semlala, arXiv: hep-ph/0211289

[ 7] \ \ N. Isgur and J. Paton: Phys. Lett. \textbf{B124 }(1983) 247, Phys.
Rev. \textbf{D31 }(1985) 2910; N. Isgur, R. Kokosky and J. Paton: Phys. Rev.
Lett. \textbf{54 }(1985) 869; F. E. Close and P. R. Page: Nucl. Phys. 
\textbf{B433 }(1995) 233, Phys. Rev. \textbf{D52 }(1995) 1706; T. Barnes, F.
E. Close and E. S. Swanson: Phys. Rev. \textbf{D5}2 (1995) 5242

[ 8] \ \ S. R. Cotanch and F. J. Llanes-Estrada: hep-ph/0008337,
hep-ph/0009191

[ 9] \ \ H. Y. Jin, J. C. K\"{o}rner and T. G. Steele; arXiv:
hep-ph/0211304v1; F. de Viron and J. Govaerts; Phys. Rev. Lett. \textbf{53}
(1984) 869

[10] \ \ T. Barnes: Caltech Ph.D. thesis (1977), unpublished, Nucl. Phys. 
\textbf{B158}, 171 (1979); T. Barnes and F. E. Close: Phys. Lett. \textbf{%
116B}, 365 (1982); M. Chanowitz and S. R. Sharpe: Nucl. Phys. \textbf{B222},
211 (1983); T. Barnes, F. E. Close and F. deViron: Nucl. Phys. \textbf{B224}%
, 241 (1983); M. Flensburg, C. Peterson and L. Skold: Z. Phys. \textbf{C22},
293 (1984); P. Hasenfratz, R. R. Horgan, J. Kuti and J.-M. Richard: Phys.
Lett. \textbf{95B}, 299 (1980)

[11] \ \ C. Michael: hep-lat/9209014, hep-ph/9810415, hep-lat/9904013,
hep-ph/0009115, hep-ph/0101287; C. Michael and S. Perantonis: Nucl. Phys. 
\textbf{B347 }(1990) 854; P. Boyle, P. Lacock, C. Michael and P. Rowland:
Phys.Rev.\ \textbf{D54 }(1996) 6997, Nucl. Phys. Proc. Supp. \textbf{63 }%
(1998) 203; C. Bernard et \textit{al}: Phys. Rev. \textbf{D56 }(1997) 7039

[12] \ \ E.S. Swanson, preprint CTP\#2047 (1992), Annals Phys.\textbf{220}%
(1992)73

[13] \ \ A.U. Badalyan, preprint LNF-91/017 (R)

[14] \ \ A. Donnachie and Yu.S. Kalashnikova, Phys.Rev\textbf{D60}%
(1999)114011

[15] \ \ C.J. Burden and M.A. Pichowsky, arXiv: hep-ph/0206161

[16] \ \ A.R. Dzierba and al, arXiv: hep-ex/0304002

[17] \ \ D.V. Bugg, private communication

\begin{equation*}
\begin{tabular}{|c|c|c|c|}
\hline
Experiment & Mass (MeV / c$^{2}$) & width (MeV / c$^{2}$) & decay mode \\ 
\hline
BNL & 1370$\pm $16$\pm _{30}^{50}$ & 385$\pm $40$\pm _{105}^{65}$ & $\eta
\pi $ \\ \hline
BNL & 1359$\pm _{14}^{16}\pm _{24}^{10}$ & 314$\pm _{29}^{31}\pm _{66}^{9}$
& $\eta \pi $ \\ \hline
CBar & 1400$\pm $20$\pm $20 & 310$\pm $50$\pm _{30}^{50}$ & $\eta \pi $ \\ 
\hline
CBar & 1360$\pm $25 & 220$\pm 90$ & $\eta \pi $ \\ \hline
CBar & $\thicksim $1400 & $\thicksim $400 & $\eta \pi $ \\ \hline
\end{tabular}%
\end{equation*}

\begin{center}
\bigskip

Table 1: detailed data for resonance $\pi _{1}(1400)^{[1]}$
\end{center}

\end{document}